\documentclass[review]{elsarticle}

\usepackage{lineno,hyperref}
\modulolinenumbers[0]

\journal{Journal of \LaTeX\ Templates}

\usepackage[american]{babel}

\usepackage{amsmath,amssymb,graphicx}
\usepackage{romannum,scalerel,stackengine}
\usepackage{tabu}
\usepackage{multirow}
\usepackage{stmaryrd}

\RequirePackage{tikz}

\newcommand{\paren}[1]{\left({#1}\right)}
\newcommand{\sqpr}[1]{\left[{#1}\right]}
\newcommand{\dydx}[2]{\frac{\mathrm{d} #1}{\mathrm{d} #2}}

\newcommand{\sech}{\mathrm{sech}}

\newcommand{\dint}{\mathrm{d}}
\newcommand{\dt}{\mathrm{d}t}
\newcommand{\pypx}[2]{\frac{\partial #1}{\partial #2}}
\newcommand{\pynpx}[3]{\frac{\partial^{#3} #1}{\partial {#2}^{#3}}}
\newcommand{\pyspx}[2]{{\partial #1}/{\partial #2}}

\newcommand{\vecd}[1]{\mathbf{#1}}

\renewcommand{\frac}[2]{\dfrac{#1}{#2}}

\newcommand{\abs}[1]{\left|{#1}\right|}

\newcommand{\quoeq}[1]{Eq.~(\ref{#1})}

\newcommand{\enavg}[1]{\left\langle{#1}\right\rangle}
\newcommand{\ppypx}[2]{\frac{\partial^2 #1}{\partial {#2}^2}}
\newcommand{\omg}{\omega_g}
\newcommand{\omoff}{\omega_{\mathrm{off}}}

\newcommand{\curbra}[1]{\left\{ #1 \right\}}
\newcommand{\quoeqq}[2]{Eqs.~({\ref{#1}}) and~({\ref{#2}})}

\renewcommand{\vecd}[1]{\mathbf{#1}}

\newcommand{\lved}{\hat{\vecd{e}}}

\nolinenumbers

\begin{document}

\begin{frontmatter}

\title{Efficiently Modeling the Noise Performance of Short-Pulse Lasers with a Computational Implementation of Dynamical Methods}

\author{Shaokang Wang,$^*$ Thomas F. Carruthers, and Curtis R. Menyuk}
\address{Department of Computer Science and Electrical Engineering, \\ University of Maryland, Baltimore County, 1000 Hilltop Circle, Baltimore, MD, 21250}
\cortext[mycorrespondingauthor]{swan1@umbc.edu}

\begin{abstract}
Lowering the noise level of short pulse lasers has been a long-standing effort for decades. 
Modeling the noise performance plays a crucial role in isolating the noise sources and reducing them. 
Modeling to date has either used analytical or semi-analytical implementation of dynamical methods or Monte Carlo simulations. 
The former approach is too simplified to accurately assess the noise performance in real laser systems, while the latter approach is too computationally slow to optimize the performance as parameters vary over a wide range.
Here, we describe a computational implementation of dynamical methods that allows us to determine the noise performance of a passively modelocked laser within minutes on a desktop computer and is faster than Monte Carlo methods by a factor on the order of $10^3$.
We apply this method to characterize a laser that is locked using a fast saturable absorber---for example, a fiber-based nonlinear polarization rotation---and a laser that is locked using a slow saturable absorber---for example, a semiconductor saturable absorbing mirror. 
\end{abstract}


\end{frontmatter}

\section{Introduction}

The search for robust, low-noise short-pulse laser sources has attracted significant attention during the last two decades~\cite{haus902165,2002ultrafast,Diddams:10,Newbury:2011,Schibli:04}.
These sources have applications to basic physics, astrophysics, environmental sciences, medicine, metrology, and many other fields~\cite{Malinauskas:2016,fenstermacher2017high,Fermann2009,ye2006femtosecond}. 
The most challenging design problems for any resonator---and particularly for short pulse lasers---usually include: (1) finding a region in the laser's adjustable parameter space where the laser operates stably, (2) optimizing the pulse profile within that region, and (3) lowering the noise levels.
Typical design objectives include optimizing the pulse profiles---such as increasing the pulse energy and decreasing the pulse duration---and lowering noise sources, which might include relative intensity noise (RIN), frequency drift, and the pulse timing and phase jitter~\cite{haus206583, Paschotta:10,Kim2014}.  
Adjustable parameters will typically include the cavity length,
the pump power, and the amplifier gain, which may be a function of not only the pump power, but also of the pump wavelength, the material, and the geometry of the gain media~\cite{Giles1991}.

In this paper, we focus on short-pulse lasers, and more particularly on passively modelocked lasers, which are the short-pulse lasers that produce  the shortest pulses. 
However, the computational method that we describe here can be applied to any resonator that can be mathematically modeled at the lowest order by the nonlinear Schr\"odinger equation, including microresonators~\cite{Chembo2013}. 

The Haus modelocking equation (HME) is the simplest and most widely-used model for passively modelocked lasers~\cite{haus902165,2002ultrafast,kartner2004few}. 
It may be written
\begin{align}\label{eq:hme}
\begin{split}\displaystyle
\pypx{u}{T}=&\bigg[-i\phi + t_s\pypx{u}{t}-\frac{l}{2}+\frac{g(\abs{u})}{2} \bigg(1 + i{\omoff\over\omg}\pypx{}{t} +{1\over2\omg^2}\ppypx{}{t}\bigg) - \frac{i\beta''}{2}\pynpx{}{t}{2} \\
&+ i\gamma|u|^2 + f_\mathrm{sa}(u) \Big]u + s(t,T), 
\end{split}
\end{align}
where $u(t,T)$ is the complex field envelope, $t$ is the fast retarded time, $T$ is the slow time of propagation normalized to the round trip time $T_R$, $\phi$ is the phase rotation per unit length per round trip, $t_s$ is the shift in $t$ of the pulse centroid $t_c=\int t'|u|^2\dint t'/\int |u|^2\dint t'$ per round trip, $l$ is the linear loss coefficient, $g(|u|)$ is the saturated gain, $\beta''$ is the group velocity dispersion coefficient, $\gamma$ is the Kerr coefficient, $\omega_g$ is the gain bandwidth, $f_\mathrm{sa}(u)$ is the saturable absorption, and $s(t,T)$ is the noise source. 
Here, we are effectively assuming a parabolic
gain model whose peak may have an offset with respect to the central
frequency $\omoff$ and has a gain bandwidth $\omega_g$,
It is common in studies of the HME to set $\phi=0$~\cite{haus902165,kutz:06,Haus:1991,Hofer1992,Kapitula:02}, in which case the phase of the pulse solution rotates at a constant rate as a function of $T$. 
In computational work, it is more useful to ensure that the solution is stationary, in which case $\phi\ne 0$. 

In the HME, it is assumed that the gain response of the medium is much longer than the round trip time $T_R$, in which case the saturable gain becomes
\begin{align}\label{eq:gain_sat}
	g(\abs{u})=\frac{g_0}{1+P_\mathrm{av}(\abs{u})/P_\mathrm{sat}},
\end{align}
where $g_0$ is the unsaturated gain, $P_{\mathrm{av}}(\abs{u})$ is the average power, 
and $P_\mathrm{sat}$ is the saturation power. 
We may write $P_\mathrm{av}(\abs{u})=\int_{-T_R/2}^{T_R/2} |u(t,T)|^2 \dt/T_R$.
In the HME, the saturable absorption is fast, i.e., the response to the incoming pulse is instantaneous, so that
\begin{align}\label{eq:fsa_hme}
	f_\mathrm{sa}(u) = \delta |u|^2,
\end{align}
where $\delta$ is the fast saturable absorption coefficient. 

When the noise term $s(t,T)$ is neglected in Eq.~(\ref{eq:hme}), and we assume that the parameters satisfy the special relations
\begin{align}\label{eq:ansatz}
\begin{split}
\frac{\delta\beta''}{\gamma} &= {g(|u_0|)\over2\omg^2}, \\
t_s&=\beta''\omoff\omega_g, \\
g(|u_0|)-l & = -{g(|u_0|)\over2\omg^2}(\gamma A_0^2-\beta''\omega_0^2 ),
\end{split}
\end{align}
then we find that Eq.~(\ref{eq:hme}) has the stationary solution~\cite{haus902165},
\begin{align}\label{eq:soliton}
\begin{split}
u_0(t)&=A_0\sech(t/\tau_0)\exp\sqpr{-i\omega_0 (t-t_0) + i\theta_0},\\
\omega_0 &= \omoff\omega_g, \\
   \phi& = {1\over2}(\gamma A_0^2-\beta''\omega_0^2), \\
	\tau_0 & = \sqrt{|\beta''|/\gamma}/A_0,
\end{split}
\end{align}
where $A_0>0$ determines both the amplitude and the duration of the stationary pulse, while $t_0$ and $\theta_0$ are the initial pulse centroid in $t$ and the initial optical phase. 
Given the special choice of parameters in \quoeq{eq:ansatz}, soliton perturbation theory~\cite{kaup1990} can be applied to the HME to determine the stability of the stationary solution~\cite{haus206583,Kapitula:02,Kutz2008,Kartner1996}. 
In addition, with the same parameter choice, the HME can be reduced to two pairs of Gordon processes that describe the propagation dynamics of the pulse energy, phase, frequency, and central time, from which the phase jitter, timing jitter, frequency jitter, and energy fluctuation can be calculated analytically~\cite{haus206583,Paschotta2001}. 
These analytical results have been widely used to estimate the noise performance of passively modelocked laser systems. 

There are two difficulties with this approach. 
The first is that the expression for $f_\mathrm{sa}(u)$ in \quoeq{eq:fsa_hme} is too simple to be realistic, and it predicts that the pulse solution is only stable in a small region in the parameter space, which is contrary to experimental results~\cite{Kapitula:02,Renninger2008,Cundiff2003}. 
More complex models that predict larger regions of stability and that better match the experiments have been studied~\cite{haus:fastabsorber,Paschotta:2004,Ablowitz:11,Chen:92,leblond2002, komarov2005, kamarov-quintic2005,Wang2016oe}. 
However, with the exception of the work in~\cite{Wang2016oe}, all this work relies on solving \quoeq{eq:hme} using evolutionary methods, which can be computationally inefficient and can lead to ambiguous results. 
Second, even given the expression for $f_\mathrm{sa}(u)$, there is no reason to expect that the special parameter relation in \quoeq{eq:ansatz} is valid. 
In fact, short-pulse lasers vary widely---using different types of gain media, saturable absorbers, and cavity designs. 
There is a need for computational tools that are sufficiently powerful to be able to cope with the broad range of short-pulse laser designs. 

Typical theoretical studies solve the evolution equations starting from computational noise or some other initial conditions and allow the solution to evolve until it either settles down to a stationary or periodically-stationary state or fails to settle down after a long evolution time~\cite{Soto-Crespo:96,akhmediev10400829}. 
This approach can be ambiguous, since it is often not clear how long it is necessary to wait for a pulse to settle down, and the computation time required to evolve to steady state approaches infinity in principle when the system parameters approach a stability boundary.
In prior work, we developed boundary tracking algorithms that are based on dynamical systems theory.
These algorithms are a set of computational methods that allow one to rapidly obtain the pulse profile and determine the regions of stable operation in a large parameter space~\cite{Wang2016oe,Wang:2014}. 
We previously referred these methods as ``spectral methods'' in~\cite{Menyuk:2016}. 
Here, we use the name ``dynamical methods'' to avoid possible confusion when evaluating the Fourier spectrum using this method.  

Despite the importance of characterizing the noise in short-pulse lasers, there have been relatively few computational studies of their noise performance. 
The computational studies that have been carried out use Monte Carlo simulations in which the evolution equations are repeatedly solved with different noise realizations~\cite{Paschotta2001,Werner1990,HESS1996,Qin2017}. 
Convergence of this procedure is slow, and it is too computationally intensive to be used for systematic optimization. 

In this paper, we extend the work in~\cite{Menyuk:2016} to study the noise performance of short-pulse lasers using dynamical methods. 
Here, for the first time, we describe in detail the computational procedure and quantitatively compare the computational performance of our dynamical method to Monte Carlo simulations. 

The remainder of this article is organized as follows. 
We present a general description of the system equations and the dynamical method in Sec.~\ref{sec:dynamical}. 
We present our computational efficiency tests in Sec.~\ref{sec:results}.
We conclude this article in Sec.~\ref{sec:conclude}. 

\section{The Dynamical Method\label{sec:dynamical}}

In this section, we describe the framework of the dynamical method. 

In the laser systems that we are considering, the evolution of the pulse envelope can be described by a nonlinear equation that has the form
\begin{align}\label{eq:master0}
	\pypx{u(t,T)}{T} = \hat{F}\sqpr{u(t,T),u^*(t,T)} + s(t,T),
\end{align}
where $\hat{F}(u,u^*)$ is a nonlinear function of the wave envelope $u$ and its complex conjugate $u^*$. 
In nearly all cases, the variable $u^*$ appears with one power less than $u$ in each term of $F$. 
That is the case for \quoeq{eq:hme} as well as for the models of fast saturable absorption that were considered in~\cite{Wang2016oe}. 
It is also implicitly the case for the model of a slow saturable absorber---such as a semiconductor saturable absorbing mirror (SESAM)---that we will consider. 
In this model, we have
\begin{align}\label{eq:sesam}
	f_\mathrm{sa}(u)=-\frac{\rho}{2}nu
\end{align}
in \quoeq{eq:hme}, where $\rho$ denotes the saturable loss coefficient, and $n(t,T)$ is the fraction of the population in the lower level of a two-level system and is given by the solution of the equation
\begin{align}\label{eq:n_sesam}
\pypx{n(t)}{t}=\frac{1-n}{T_A}-\frac{\abs{u(t)}^2}{w_A}n, \quad n\paren{-\frac{T_R}{2}}=0,
\end{align}
where $T_A$ and $w_A$ denote the response time and the saturation energy of the absorber. 

We assume here that a stationary solution to \quoeq{eq:master0} in the absence of noise $u_0(t)$ has been found. 
We previously described computational procedures that allow us to rapidly find stationary solutions as system parameters vary and determine their stability~\cite{Wang:2014}. 
To determine the stability, it is necessary to consider an extended system. 
Writing the complex conjugate equation of \quoeq{eq:master0} as $\pyspx{u^*}{T}=F^*$, we may write the linearized equation
\begin{align}\label{eq:ddudt}
	\pypx{\Delta \vecd{u}}{T} = \mathsf{L}\Delta \vecd{u} + \vecd{s},
\end{align}
where 
\begin{align}
	\Delta\vecd{u} = \begin{bmatrix}
	\Delta u \\ \Delta \bar{u}
	\end{bmatrix}, \quad
	\mathsf{L} = \begin{bmatrix}
	\mathsf{L}_{11} & \mathsf{L}_{12} \\ \mathsf{L}_{21} & \mathsf{L}_{22}
	\end{bmatrix}, \quad
	\vecd{s} = \begin{bmatrix}
	s \\ s^*
	\end{bmatrix},
\end{align}
where $\mathsf{L}_{11}=\delta {F}/{\delta u}$, $\mathsf{L}_{12}=\delta {F}/{\delta u^*}$, $\mathsf{L}_{21}=\delta {F^*}/{\delta u}$, and $\mathsf{L}_{22}=\delta {F^*}/{\delta u^*}$ are functional derivatives. 

We see that if $\Delta \bar{u}=\Delta u^*$ at any time $T$, then $\Delta \bar{u}=\Delta u^*$ at all times $T$. We next consider the spectrum of the operator $\mathsf{L}$ that is given by solving the eigenvalue equation
\begin{align}\label{eq:eigen-problem}
\mathsf{L}\Delta\vecd{u} = \lambda\Delta\vecd{u}.
\end{align}
If any eigenvalue has a positive real part, then the system is unstable. 

In any practical laser system, the noise $s(t,T)$ is a small perturbation. 
Indeed, it is typically so small that it is necessary to artificially increase it in order to obtain reliable results from Monte Carlo simulations~\cite{Paschotta:2004}. 
The essence of our method is that the amplitudes of the spectral components---the solutions to \quoeq{eq:eigen-problem}---obey simple Langevin equations that can be solved for all times $T$. 
The means and variances of these amplitudes can then easily be found. 
After expanding the statistical quantities of interest such as the phase jitter, the timing jitter, and the energy fluctuation as a linear sum of these amplitudes, we obtain the means and variances of these statistical quantities.

\subsection{Descretization}

When we descretize the time domain $t$ for computation, we use an evenly spaced grid of $N$ points in $t$, whose spacing we denote as $\Delta t$, where $\Delta t=T_w/N$ and $T_w$ is the duration of the computational time window. 

Issues related to choosing $\Delta t$ and $N$ as well as discretizing the operator $\mathsf{L}$ to ensure the accuracy of the solution have been discussed in~\cite{Wang:2014}. 
Here, in order to ensure reasonable accuracy, we choose $T_w$ so that it is approximately 100 times the duration of the modelocked pulse, and we choose $N\ge1024$. 
We always choose $T_w$ and $N$ sufficiently large so that the visible impact on any plotted result is negligible. 

In analytical studies of the stability and noise performance of passively modelocked lasers, it is usual to choose an infinite domain in the fast time $t$, in which case the spectrum of $\mathsf{L}$ has both continuous components (essential spectrum) as well as discrete components (point spectrum)~\cite{kaup1990, Wang:2014, Kapitula03theevans}. 
In real-world lasers, the actual domain is periodic in the round trip time $T_R$, and in computational work, it is usual to study a time domain $T_w$ that is small compared to $T_R$, so that $T_w\ll T_R$. 
As a consequence, the computational problem only has a point spectrum. 

Once the system has been discretized, both $\Delta u(t)$ and $\Delta \bar{u}(t)$ become $N$-dimensional vectors in which $\Delta {u}_l=u(t_l)$ and $\Delta \bar{{u}}_l=\bar{u}(t_l)$, $l=1,2,\cdots,N$. 
The vector $\Delta \vecd{u}$ in \quoeq{eq:ddudt} becomes a $2N$-dimensional vector $\Delta \vecd{u}$ in which the first $N$ elements correspond to $\Delta u_l$, $l=1,2,\cdots,N$ and the last $N$ elements correspond to $\Delta \bar{u}_l$, $l=1,2,\cdots,N$, i.e., $\Delta\vecd{u}=[\Delta u_1,\Delta u_2,\cdots,\Delta u_N,\Delta \bar{u}_1,\Delta \bar{u}_2,\cdots,\Delta \bar{u}_N]^T$, where $T$ denotes the transpose.
The operator $\mathsf{L}$ becomes a $2N\times2N$ matrix~\cite{Wang:2014}.

\subsection{Spectral Decomposition}

We will denote a set of independent eigenvectors as $\vecd{e}_j = [e_j,\bar{e}_j]^T$, where $T$ denotes the transpose and $e_{jl} = e_j(t_l)$ and $\bar{e}_{jl} = \bar{e}_j(t_l)$, so that each eigenvector $\vecd{e}_j$ is a $2N$-dimensional vector. 
In all the laser problems that we have considered, the set of eigenvectors $\vecd{e}_j$ is complete, i.e., there are $2N$ independent eigenvectors, which span the $2N$-dimensional complex vector space upon which $\mathsf{L}$ operates~\cite{trefethen1997numerical}, so that we may decompose any $\Delta {\vecd{u}}$ as 
\begin{align}\label{eq:define-cj}
	\Delta \vecd{u} = \sum_{j=1}^{2N} c_j\vecd{e}_j,
\end{align}
where the $c_j$ are complex constants. 
We find that if $\lambda_j$ is an eigenvalue, then so is $\lambda_j'=\lambda_j^*$ and if $\vecd{e}_j = [e_j,\bar{e}_j]^T$, then the eigenvector corresponding to $\lambda_j'=\lambda_j^*$ is given by $\vecd{e}_j' = [\bar{e}_j^*,{e}_j^*]^T$~\cite{Menyuk:2016}. 
In general $\bar{e}_j\ne e_j^*$. 
However, when $\lambda_j$ is real, then we find $\bar{e}_j= e_j^*$. 

In order to find the $c_j$, given $\Delta\vecd{u}$, we must define an inner product. 
For any two given vectors $\vecd{p}$ and $\vecd{q}$ in the $2N$-dimensional space,
the natural inner product becomes
\begin{align}\label{eq:innerprod}
\sum_{j=0}^{N}\paren{p_j^*q_j+\bar{p}_j\bar{q}_j^*}\Delta t = \vecd{p}^H\vecd{q}\Delta t,
\end{align}
where $\vecd{p}^H$ is a $2N$-dimensional row vector whose elements are complex conjugates of the vector $\vecd{p}$. 

We will denote the dual eigenvectors of the matrix $\mathsf{L}$ as $\lved_j$. 
These are equal to the eigenvectors of $\mathsf{L}^\dagger$, the complex conjugate transpose of $\mathsf{L}$. 
The dual eigenvectors are normalized so that 
\begin{align}\label{eq:bi-orth}
\hat{\vecd{e}}_j^H{\vecd{e}}_k\Delta t = \delta_{jk},
\end{align}
where $\delta_{jk}$ is the Kr\"onecker delta-function. 
We now find that 
\begin{align}
	c_j = \lved_j^H \Delta\vecd{u}.	
\end{align}
Since $\mathsf{L}\ne\mathsf{L}^\dagger$, so that $\mathsf{L}$ is not self-adjoint, it is {\it NOT} generally the case that ${\hat{\vecd{e}}_j}^H{\hat{\vecd{e}}_k}\Delta t = \delta_{jk}$.

\subsection{Noise Evolution}

In this paper, we will consider white noise sources for which 
\begin{align}\label{eq:defnoise}
	\enavg{s(t,T)s^*(t',T')}=D\delta(t-t')\delta(T-T'),
\end{align}
where $\enavg{\cdot}$ denotes the emsemble average, and $D$ is the diffusion coefficient. 
We also have $\enavg{s(t,T)s(t',T')}=\enavg{s^*(t,T)s^*(t',T')}=0$. 
More complex noise source can be built up using \quoeq{eq:defnoise} as a starting point~\cite{Docherty2013}. 
After discretization in $t$, \quoeq{eq:defnoise} becomes 
\begin{align}\label{eq:defnoise-disc}
\enavg{s_l(T)s_m^*(T')}=\enavg{s(t_l,T)s^*(t_m,T')}=\frac{D}{\Delta t}\delta_{lm}\delta(T-T'),
\end{align}
where $s_l = s(t_l)$, and the $2N$-dimensional vector $\vecd{s}$ becomes $\vecd{s}=[s,\bar{s}]^T$, where $\bar{s}_l = s_l^*$.

After discretization, we can write the $2N$-dimensional vector $\vecd{s}$ at any slow time $T$ as
\begin{align}
	\vecd{s}(T) = \sum_{j=1}^{2N}s_j(T)\vecd{e}_j,
\end{align}
so that ${s}_j(T) = \lved_j^H\vecd{s}(T)$. 
We now find 
\begin{align}
\enavg{s_j(T)s^*_k(T')} = \paren{D\Delta t} {\lved_j}^H{\lved_k}\delta(T-T') = D_{jk}\delta(T-T'), \label{eq:cross-sjsk}
\end{align}
where we note $D_{kj} = D_{jk}^*$.

In the presence of noise, we find that the amplitudes of the spectral components of $\Delta \vecd{u}$ that are defined in \quoeq{eq:define-cj} evolve according to the simple Langevin equation
\begin{align}\label{eq:cj-evo}
\dydx{c_j}{T} =  \lambda_j c_j + s_j,
\end{align}
where we note that $\mathrm{Re}(\lambda_j)\le0$ in order for the modelocked pulse to be stable. 
Since we start from a stationary solution, we now have $\enavg{c_j(T=0)} = 0$. 

The covariances, which can be obtained by integrating \quoeq{eq:cj-evo} using the method of stochastic differential equations~\cite{Wai1996}, become
\begin{align}\label{eq:variance0}
\enavg{c_j(T)c_k^*(T)} =-\frac{D_{jk}}{\lambda_j +\lambda_k^*} \sqpr{1-e^{\paren{\lambda_j+\lambda_{k}^*} T_2}},
\end{align}
where we assume that the covariances are zero at $T=0$. 
In the special case when $\lambda_j=\lambda_{k}=0$, we obtain
\begin{align}
\enavg{c_j(T)c_k^*(T)} =D_{jk}T.
\end{align}
In the long-time limit as $T\to\infty$, \quoeq{eq:variance0} becomes
\begin{align}\label{eq:variance1}
\enavg{c_j(T)c_k^*(T)} = -\frac{D_{jk}}{\lambda_j +\lambda_k^*}.
\end{align}

The corresponding two-time correlation function as $T\to\infty$ is giving by~\cite{Wai1996}
\begin{align}\label{eq:cj-crosscorr}
R_{jk}(\tau) & = -\frac{D_{jk}}{\lambda_j +\lambda_k^*}\sqpr{ e^{\lambda_k^*\tau}\Theta(\tau)+ e^{-\lambda_j\tau}\Theta(-\tau)},
\end{align}
where $\Theta(\tau)$ is the Heaviside step function that equals zero when $\tau<0$, 1/2 when $\tau=0$, and 1 when $\tau>0$. 
The corresponding power spectral density is given by the Fourier transform of $R_{jk}(\tau)$,
\begin{align}\label{eq:cjk-psd}
S_{jk}(f) = \frac{D_{jk}}{\paren{\lambda_j-2i\pi f}\paren{\lambda_k^*+2i\pi f}},
\end{align}
Using Eqs.~(\ref{eq:variance0})--(\ref{eq:variance1}), it is possible to compute quantities of statistical interest such as the timing jitter and the phase jitter.
Using \quoeqq{eq:cj-crosscorr}{eq:cjk-psd} it is then possible to calculate the power spectral densities of these quantities. 

\subsection{Noise Impact on Statistical Quantities of Interest\label{sec:noise}}

{Given a statistical of interest, $\Delta x(T)$, we begin by writing it as an inner product of an appropriate vector $\vecd{h}_x$ and the perturbation $\Delta \vecd{u}(T)$,}
\begin{align}\label{eq:def-h}
\Delta x(T) = \vecd{h}_x^H \Delta\vecd{u}(T)\Delta t,
\end{align}

Some examples follow:
\begin{enumerate}
	\item Energy jitter $\Delta w(T)$: 
	
	The energy jitter is given by
	\begin{align}
	\begin{split}
	& \Delta w (T) = \int_{-T_R/2}^{T_R/2}\dint t \sqpr{|u(t,T)|^2-|u_0(t)|^2} \\
	&  = \int_{-T_R/2}^{T_R/2}\dint t \sqpr{u_0(t)\Delta u^*(t,T) + u_0^*(t)\Delta u(t,T)}, 
	\end{split} \nonumber
	\end{align}
which becomes after discretization
\begin{align}\label{eq:def-dw}
\begin{split}
	\Delta w(T) & = \sum_{l=1}^{N} \Delta t\sqpr{u_0(t_l)\Delta u^*(t_l,T) + u_0^*(t_l)\Delta u(t_l,T)}\\
	&=\vecd{h}_w^H{\Delta\vecd{u}(T)}\Delta t
\end{split}
\end{align}		
where $\vecd{h}_w = [u_0,u_0^*]^T$. 	
	\item Frequency jitter $\Delta f_c(T)$~\cite{Grigoryan1999}:
		
	We can calculate the change in the central frequency as
	\begin{align}
	\begin{split}
	\Delta f_c(T) = \frac{1}{2iw_0} \int_{-T_R/2}^{T_R/2}\dint t 
	\sqpr{ \frac{\partial u_0^*}{\partial t}\Delta u (t,T) - \frac{\partial u_0}{\partial t} \Delta u^*(t,T)}, \\
	\end{split}
	\end{align}
	which after discretization becomes
	\begin{align}\label{eq:def-df}
	\begin{split}
	& \Delta f_c(T) = \vecd{h}_{f_c}^H{\Delta\vecd{u}(T)}\Delta t,
	\end{split}
	\end{align}
	where $\vecd{h}_{f_c} = ({i}/{w_0})\sqpr{\mathsf{D}_t u_0,\mathsf{D}_t {u}^*_0}^T$, where $\mathsf{D}_t$ is a first-order differentiation matrix, which we obtain by using the Fourier transform to compute $u_0$ in the frequency domain, multiplying by the frequency, and then computing the inverse Fourier transform~\cite{Weideman2000}. 
	
	\item Timing and phase jitter:
	
	The central time of a modelocked pulse is given by 
	\begin{align}
	\Delta t_c & = {1\over w_0}\int_{-T_R/2}^{T_R/2}\dint t\  t\big[u_0^*(t)\Delta u(t,T)
	+ u_0(t) \Delta u^*(t,T)\big],
	\end{align}
	which after discretization becomes
	\begin{align}
		\Delta t_c = \vecd{h}_{t}^H{\Delta\vecd{u}(T)}\Delta t,
	\end{align}
	where $\vecd{h}_{t} = (1/w_0)[tu_0,tu_0^*]^T$.
	
	From the timing jitter, we can define a phase jitter,
	\begin{align}
	\Delta \psi = 2\pi\Delta t_c/T_R,
	\end{align}
	which corresponds to the phase jitter that is observed at radio frequencies after an optical signal is detected in a photodetector. 
	In most experimental work, this quantity is simply referred to as the phase jitter. 
	Paschotta~\cite{Paschotta:2004} refers to it as the timing phase jitter to avoid confusion with the optical phase jitter, and we will do the same. 
\end{enumerate}

In general, for any vector $\vecd{h}_x$, we can write 
\begin{align}\label{eq:span-h}
	\vecd{h}_x = \sum_{j=1}^{2N} h_{xj} \hat{\vecd{e}}_j,
\end{align}
and combined with \quoeq{eq:def-h}, the corresponding statistical quantity can be written as
\begin{align}\label{eq:hT-expression}
	\Delta x(T) = \Delta t\paren{\sum_{j=1}^{2N} h_{xj} \lved_j}^H{\sum_{{k}=0}^{2N} c_k(T) \vecd{e}_k} = \sum_{j=1}^{2N} h_{xj}^*c_{j}(T),
\end{align}
where the $h_{xj}$ can be derived using
\begin{align}\label{eq:calculate-h}
h_{xj} = \vecd{e}_j^H{\vecd{h}_x}\Delta t.
\end{align}

Following \quoeqq{eq:cjk-psd}{eq:hT-expression}, we can now calculate the power spectral density of $\Delta x(T)$,
\begin{align}\label{eq:psd-h}
	S_{x}(f)= \sum_{j=1}^{2N}\sum_{{k}=1}^{2N} h^*_{xj}h_{xk} S_{{jk}}(f) = \sum_{j=1}^{2N}\sum_{{k}=1}^{2N}\frac{h_{xj}^*h_{xk}D_{jk}}{\paren{\lambda_j-2i\pi f}\paren{\lambda_k^*+2i\pi f}},
\end{align}
in which we require $h_{xl}=0$ when $\lambda_l=0$.

Defining $\delta c_j={\dint c_j}/{\dint T}$, we have
\begin{align}\label{eq:dhT-expression}
	\dydx{\Delta x}{T} = \delta x(T) = \sum_{j=0}^{2N} h_{xj}^* \delta c_j(T),
\end{align} 
which approximates the change in $\Delta x(T)$ from one round trip to the next, since all statistical quantities of interest change slowly compared to the repetition time. 
The power spectral density of $\delta x(T)$ becomes
\begin{align}\label{eq:psd-dh}
	S_{\delta x}(f) = {(2\pi f)^2S_{x}(f)}.
\end{align}
The formalism in \quoeqq{eq:psd-h}{eq:psd-dh} includes the contribution of the eigenvectors that correspond to the continuous spectrum, whose effects were neglected in~\cite{Menyuk:17}.

\section{Noise Level Evaluation and Computational Efficiency Tests\label{sec:results}}

Here, we compare the results of the Haus-Mecozzi method~\cite{haus206583}, the Monte Carlo method~\cite{Paschotta2001}, and the dynamical method that we have described in Sec.~\ref{sec:dynamical}. 
The statistical quantities that we will study are the energy jitter $\Delta w(T)= w(T)-w_0$, the frequency jitter $\Delta f_c(T) = f_c(T) - f_0$, and the timing phase jitter $\Delta t_c = t_c(T) - t_{c0}$, where $w_0$, $f_0$, and $t_{c0}$ are the unperturbed energy, central frequency, and the central time of the modelocked pulse. 
We first give a brief review of the three methods that we will compare. 
We then apply all three methods to the widely-used Haus modelocking equation (HME) and an averaged model of a SESAM fiber laser~\cite{Sinclair:14}. 
We show that the dynamical method provides significantly better agreement with the Monte Carlo method than does the Haus-Mecozzi method.
We further show that the dynamical method is several orders of magnitude more computationally efficient than the Monte Carlo method, where our metrics are the computational time and the memory (RAM) and storage usage.  

\subsection{Calculation Methods\label{sec:approaches}}
We first review the three methods we use to calculate the noise impact on the statistical quantities of interest. These are: (1) the Haus Mecozzi method, which is analytical, (2) the Monte Carlo simulation method, which repeatedly solves the evolution equations with different noise realizations, and (3) the dynamical methods that we described in Sec.~\ref{sec:dynamical}.

\subsubsection{The Haus-Mecozzi Method\label{sec:hmnm}}
The Haus modelocking equation (HME) is the simplest and most widely used model for modelocked laser systems. 
We have presented the HME in Eqs.~(\ref{eq:hme})--(\ref{eq:soliton}). 
In their analytical method, Haus and Mecozzi begin by assuming that the modelocked pulse $u_0(t)$ has a hyperbolic-secant pulse shape and---like the soliton solutions for the noinlinear Schr\"odinger equation---is completely characterized by four parameters: the pulse energy and its central time, central phase, and central frequency. 
They next apply soliton perturbation theory to calculate the phase evolution in the presence of noise, and they show that the evolution of the pulse energy fluctuation $\Delta w$, the central phase fluctuation $\Delta\theta$, the central frequency fluctuation $\Delta f_c$, and the central time fluctuation $\Delta t_c$ are governed by four stochastic differential equations~\cite{haus206583,Paschotta:2004},
\begin{align} \label{eq:odes-hmn}
	\begin{split}
		{\dint \Delta w}/{\dint T} &= r_w \Delta w + s_w, \\
		{\dint \Delta \theta}/{\dint T} &= r_\theta\Delta w + s_\theta, \\
		{\dint \Delta f_c}/{\dint T} &= r_f\Delta f_c + s_f, \\
		{\dint \Delta t_c}/{\dint T} &= r_t\Delta f_c + s_t, 
	\end{split}
\end{align}
where the growth/decay coefficients are all real quantities,  
\begin{align}
	\begin{split}
		r_w &= 2\delta A_0^2 -g_1w_0 + 2g_1 A_0^2/\paren{6\omega_g^2\tau_0}, \\
		r_\theta &= \gamma A_0^2/w_0, \\	
		r_f &= -{g_\mathrm{sat}}/\paren{3\omega_g^2\tau_0^2}, \\
		r_t &= \beta'', 
	\end{split}
\end{align}
and for which $g_1 = g_\mathrm{sat}^2/(g_0P_\mathrm{sat}T_R)$, $g_\mathrm{sat} = g(|u_0(t)|)$, and $w_0 = 2A_0^2\tau_0$ is the energy of the modelocked pulse. 
The diffusion coefficients are defined as $\enavg{s_x(T),s_x^*(T')}=D_x\delta(T-T')$ for $x=w,\theta,f,t$,
\begin{align}
	\begin{split}
		D_w &= 2w_0 D, \\
		D_\theta &= 2D(1+\pi^2/12)/\paren{3w_0}, \\
		D_f &= 2D/\paren{3w_0\tau_0^2}. \\
		D_t &= \pi^2\tau_0^2D/\paren{6w_0}, 
	\end{split}
\end{align}
where $D$ is defined in Eqs.~(\ref{eq:defnoise}) and~(\ref{eq:defnoise-disc}).
These four quantities $\Delta w$, $\Delta\theta$, $\Delta f_c$, and $\Delta t_c$ correspond to the magnitudes of the four discrete eigenmodes in the spectrum of the linearized Haus-Mecozzi model~\cite{haus206583}. 
We note that $\Delta\theta$ corresponds to the optical phase jitter, which is rarely measured. 

The stochastic differential equations in \quoeq{eq:odes-hmn} can be solved analytically. 
The variances of $\Delta w(T)$, $\Delta f_c(T)$, and $\Delta t_c(T)$ become
\begin{align}\label{eq:variances-hmn}
	\begin{split}
		\sigma_w^2(T) = & \enavg{|\Delta w(T)|^2} = -{D_w}(1-e^{2r_wT})/(2r_w) \xrightarrow{T\to\infty} -D_w/(2r_w),  \\
		\sigma_{f_c}^2(T) = & \enavg{|\Delta f_c(T)|^2} =  -{D_f}(1-e^{2r_fT})/(2r_f) \xrightarrow{T\to\infty} -D_f/(2r_f), \\
		\sigma_{t_c}^2(T) =& \enavg{|\Delta t_c(T)|^2} =  ({r_t^2}D_f/{r_f^2} + D_t)T + {2r_t^2}D_t(1-e^{r_f T})/r_f^3 \\ & - {r_f^2}D_t (1-e^{2r_fT})/({2r_f^3}) \xrightarrow{T\to\infty} D_t T + (1/3)D_f r_f^2T^3, 
	\end{split}
\end{align}
which indicates that the variances of energy and the frequency will remain constrained as $T\to\infty$, while the variance of the central time is unbounded. 
In experiments, the timing phase jitter is defined by the central time drift between two consecutive round trips~\cite{Paschotta:2004}, which we approximate as  $\delta t_c = \dint \Delta t_c/\dint T$. 

The Langevin equations that we introduced in \quoeq{eq:cj-evo} and the variances of the statistical quantities that we introduced in \quoeq{eq:variance1} effectively generalize \quoeqq{eq:odes-hmn}{eq:variances-hmn} to any modelocked pulse waveform and any governing equation that has the form of \quoeq{eq:master0}.
The power spectral densities for $\Delta w$, $\Delta f_c$, and $\Delta \psi$~\cite{haus206583,Paschotta:2004} are
\begin{align}\label{eq:psd-hmn}
	\begin{split}
		S_{w}(f) & = \frac{D_w}{r_w^2+(2\pi f)^2},\\
		S_{f_c}(f) & = \frac{D_f}{r_f^2+(2\pi f)^2}, \\
		S_{\psi}(f) & = \frac{S_{\delta t_c}(f)}{(T_Rf)^2} = \frac{r_t^2 D_f}{(T_Rf)^2\sqpr{r_f^2+(2\pi f)^2}} + \frac{D_t}{(T_Rf)^2}.
	\end{split}
\end{align}

\subsubsection{The Monte Carlo Simulation Method}

For a given set of parameters, we carry out a large number of Monte Carlo simulation runs with independent noise realizations. 
In each simulation run, we solve the laser evolution equation, \quoeq{eq:hme}, using a variant of the split-step method~\cite{Wang2013}. 
We use the local error to adjust the propagation step sizes~\cite{Sinkin:03}.
We use $N_\mathrm{mc}$ to denote the number of simulation runs, and we use $N_{R}$ to denote the number of round trips in each run. 
For a given statistical quantity $\Delta x(T)$, we obtain a time series $\Delta x[k] = \Delta x(kT_R)$, $k=1,2,\cdots,N_R$. 

We finally evaluate the power spectrum of a given time series $\Delta x[k]$ using the discrete-time Fourier transform and the ensemble average over all the runs,
\begin{align}\label{eq:psdEvaluate}
	\bar{S}_h(f) = \frac{1}{N_\mathrm{mc}N_R}\sum_{n=1}^{N_\mathrm{mc}}
	\abs{\mathrm{DTFT}\curbra{\Delta x[k]}}^2,
\end{align}
where in this study we set $N_\mathrm{mc}=600$, and $N_R=12000$.

\subsubsection{The Dynamical Method}

In Sec.~\ref{sec:dynamical}, we have described the derivation and the implementation of the dynamical method. 
	
\subsection{Application to Modelocked Systems}

We now compare the three different methods that we summarized in Sec.~\ref{sec:approaches}. 
In Secs.~\ref{sec:dynamical} and~\ref{sec:hmnm}, we formulated the dynamical method and the Haus-Mecozzi method in terms of the normalized frequency. 
In order to plot the noise spectrum in terms of the physical frequency $f_\mathrm{phys}$, we substitute 
\begin{align}
	f = f_\mathrm{phys}T_R.
\end{align}

\subsubsection{The Haus Modelocking Equation~\label{sec:hmn-compute}}

We first perform a comparison of the computational efficiency of these three methods with the HME~\cite{haus206583}, given in Eqs.~(\ref{eq:hme})--(\ref{eq:soliton}), and setting
\begin{align}\label{eq:ase}
D = g(|u_0|)h\nu_0 T_R,
\end{align}
where $h$ is Planck's constant, and $\nu_0$ is the central frequency of the optical field.
The computations are carried out using Matlab{\textregistered} on a desktop workstation, Dell{\textregistered} Precision Tower 7910 which uses an Intel{\textregistered} Xeon(R) CPU E5-2630 v4 with 10 cores. 
The system memory is 16\,GB. 
The operation system is Ubuntu 16.04 LTS. 
Matlab{\textregistered} uses about 500\,MB when it is started without running any programs.  
We use the parameters from~\cite{Paschotta:2004} and show them in Table~\ref{tab:params-hmn}.

\begin{table}[h!]
	\begin{center}
		\begin{tabular}{|c|l||c|l||c|l|}
			\hline
			Parameter & Value & Parameter & Value & Parameter & Value \\
			\hline
			$T_R$ & 10\,ns & $g_0$ & 0.603 & $\omega_g$ & 20\,T\,rad/s \\
			$\gamma$ & $1/\mathrm{MW}$ & $\nu_0$ & 282\,THz & $l$ & 0.0563 \\				
			$P_\mathrm{sat} T_R$ & 2\,nJ & $\beta''$ & $-0.003$\,ps$^2$ & $\delta$ & $0.046/\mathrm{MW}$ \\
			$w_0$ & 20\,nJ & $A_0$  & $182.5\,\sqrt{\mathrm{W}}$ & $\tau_0$ & 0.3\,ps \\			
			\hline
		\end{tabular}
		\caption{The parameters we use to evaluate the noise levels. These parameters are the same as in~\cite{Paschotta:2004}. \label{tab:params-hmn}}
	\end{center}
\end{table}

\begin{figure}[h!]
	\begin{center}
		\includegraphics[width=0.65\linewidth]{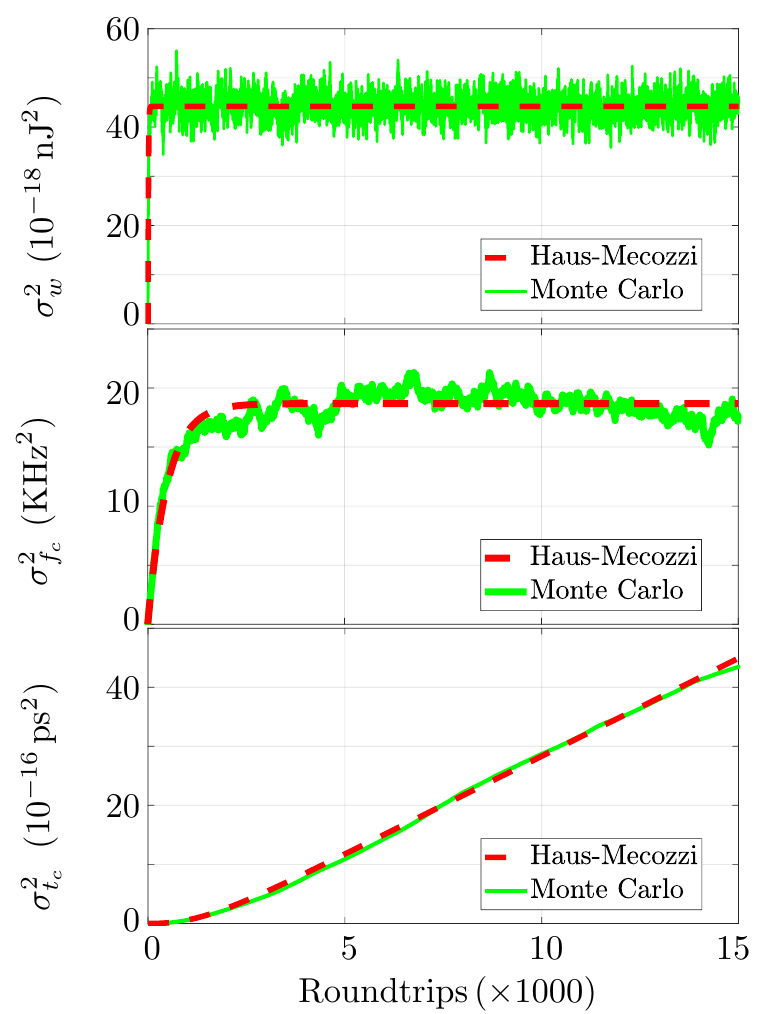}		
		\caption{Comparison between the Haus-Mecozzi and Monte Carlo methods, where $\sigma_{w}^2(T)$, $\sigma_{f_c}^2(T)$, and $\sigma_{t_c}^2(T)$ are propagation-dependent variances of the pulse energy $w$, central frequency $f_c$, and the central time $t_c$. The  results of the Haus-Mecozzi method are from \quoeq{eq:variances-hmn}. \label{fig:propagate-deltas}}
	\end{center}
\end{figure}

We propagate the laser system for 15000 round trips and we observe that the statistical properties of the noise-related quantities---the pulse energy, the central frequency, and the rate of change of the round trip time---appear stationary after 3000 round trips. 
The propagation of the variances of $\Delta w$, $\Delta f_c$, and $\Delta t_c$ are shown in Fig.~\ref{fig:propagate-deltas}.
The variances of $\Delta w$ and $\Delta f_c$ eventually reach an asymptote, while the variance of $\Delta t_c$ grows indefinitely, which agrees with \quoeq{eq:variances-hmn}.

\begin{figure}[h!]
	\begin{center}
		\includegraphics[width=0.7\linewidth]{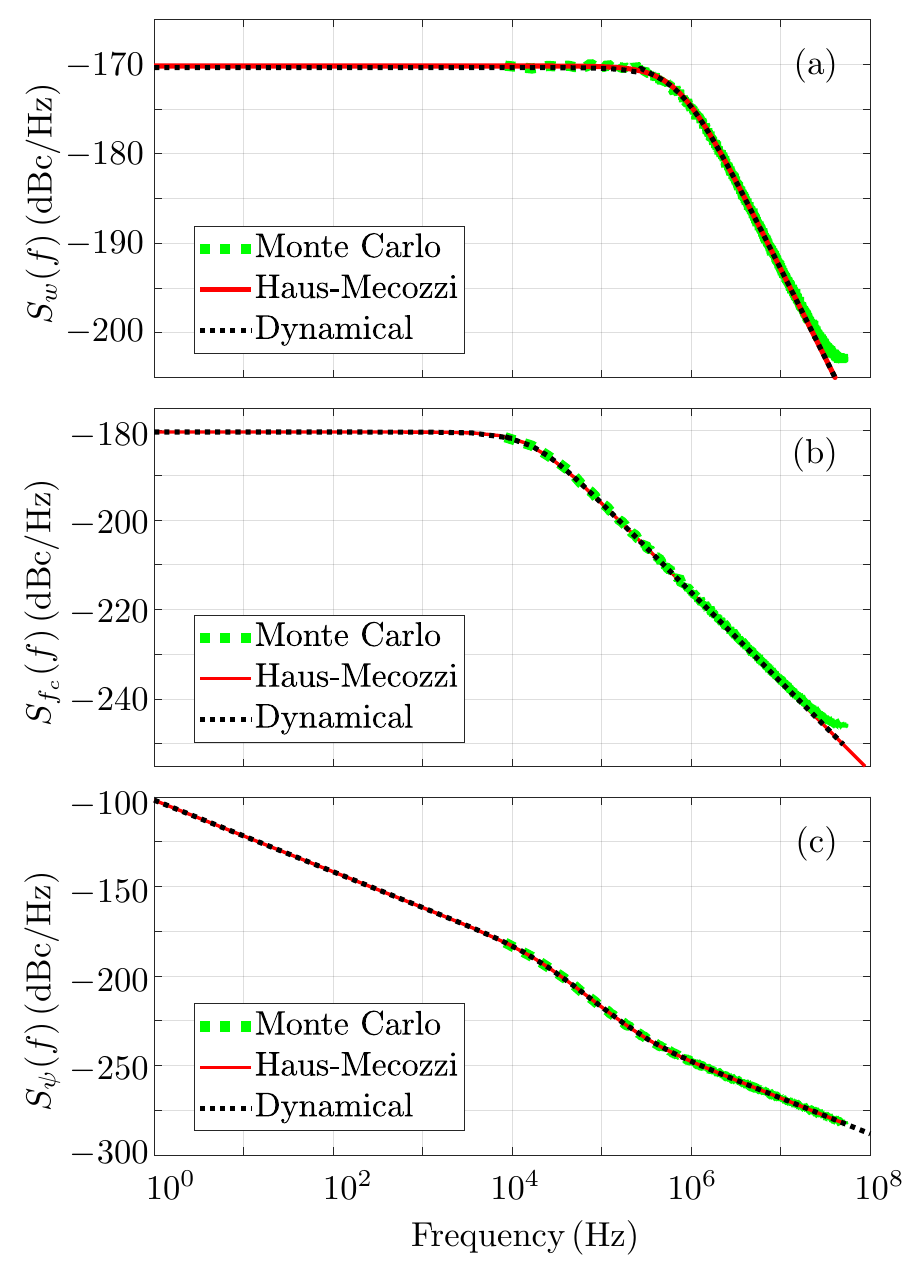}								
		\caption{The noise spectra of (a) the energy jitter, (b) the frequency jitter, and (c) the timing phase jitter that we obtain from the Monte Carlo, Haus-Mecozzi, and dynamical methods. The agreement is excellent and the results in (c) agree with Fig.~1 in~\cite{Paschotta:2004}. \label{fig:psds-hmn}}
	\end{center}
\end{figure}

In Fig.~\ref{fig:psds-hmn} we show the power spectral densities that we obtain.
All spectra are single-sided spectra~\cite{Paschotta:2004}. 
In Fig.~\ref{fig:psds-hmn}(a) we plot the energy noise as $10\log_{10}\sqpr{S_{w}(f)/w_0^2}$. 
the frequency noise as $10\log_{10}\sqpr{S_{f_c}(f)/\nu_0^2}$, and the phase noise as $10\log_{10}\sqpr{S_\psi(f)}$ which is consistent with Fig.~1 in~\cite{Paschotta:2004}.
For all three power spectral densities, the agreements of the three methods is excellent.

In Fig.~\ref{fig:psds-hmn}, we plot the spectrum from $1$\,Hz to $10^8$\,Hz. 
The Haus-Mecozzi method produces analytical predictions and thus can be used at any frequency resolution. 
The dynamical method can also be used at any frequency resolution. 
When evaluating the noise spectrum in the Monte Carlo method, we assign $N_\mathrm{mc}=600$ and $N_R = 12000$, which enables us to show the frequency range from about $8$\,kHz to 50\,kHz. 
Any increase in the frequency resolution greatly increases the computational load when using the Monte Carlo method, which imposes a practical limit on the frequency resolution that can be obtained. 

\begin{table}
	\begin{center}
		\begin{tabular}{|l|c|c|c|c|c|}
			\hline
			Method & \# of cores & Time cost & Memory usage & Storage usage \\
			\hline
			A single run & 1 & 7.8\,s & 535\,MB & 1.1 MB\\
			\hline
			600 runs & 6 & 784\,s & 2.87\,GB & 245.8 MB\\
			\hline 
			Dynamical & 1 & $<3$\,sec & 967\,MB & 141.5\,MB\\
			\hline
		\end{tabular}
		\caption{Comparison of the computational efficiency of the Monte Carlo and dynamical methods for evaluating the noise performance of the Haus modelocking equation. We integrate the system for 15000 round trips on each simulation run of the Monte Carlo method. The tests are coded in Matlab{\textregistered} which have a memory overhead of 500\,MB that is included in the memory usage. \label{tab:costs-hmn}}
	\end{center}
\end{table}

The time and memory cost performances of the Monte Carlo method and the dynamical method are summarized in Table~\ref{tab:costs-hmn}. 
We achieve a good agreement with the Haus-Mecozzi and the dynamical methods when we use the Monte Carlo method with 600 simulations. 
The total CPU time cost is about ($784\times6=4704$)\,sec, which is about 1 hour and 18 min. 
The memory usage per core ($2870/6\approx478$)\,MB, which is less than that for a single run (535\,MB) because the overhead of parallel computing is spread when more nodes are used. 
More memory might be required if a finer discretization of $u(t,T)$ in both $t$ and $T$ is needed. 
The storage usage is low (less than 1\,GB) in the Monte Carlo simulations since we only save the pulse parameters, $f_c$, $w$, and $t_c$, instead of saving the pulse profile for each iteration. 

The dynamical method has a far greater computational efficiency than does the Monte Carlo method. 
The dynamical method is able to cover a larger frequency range than does the Monte Carlo method in less than 3\,sec of computational time. 
In the example shown here, we calculated 80 frequencies from 1\,Hz to 80\,Hz.
The dynamical method uses more memory in a single core than does the Monte Carlo method, but the total memory use is still less than 1\,GB.

\subsubsection{The SESAM Laser}

Next, we consider a case when there is no known analytical solution. 
Here, we model a laser with a semiconductor saturable absorption mirror (SESAM), in which saturable absorber responds slowly compared to the time duration of the modelocked pulse~\cite{Kartner1996}
Typical time scales are picoseconds for the response time of the SESAM and 100--200\,femtoseconds for the pulse duration, as we show in Table~\ref{tab:params-sesam}~\cite{Wang:2017}.
The central wavelength of the output pulse is $1564$\,nm. 
The system can be described using Eqs.~(\ref{eq:hme}),~(\ref{eq:gain_sat}),~(\ref{eq:sesam}),~(\ref{eq:n_sesam}) and~(\ref{eq:ase}). 

\begin{table}
	\begin{center}
		\begin{tabular}{|c|l||c|l||c|l|}
			\hline
			Parameter & Value & Parameter & Value & Parameter & Value \\
			\hline
			$T_R$ & 3.33\,ns &  $w_A$ & 157 pJ & $P_\mathrm{sat}$ & 9.01\,mW \\
			$g_0$ & 7.74 &  $\rho$ & 0.0726 & $\beta''$  & $-0.0144$ ps$^{2}$ \\
			$\omega_g$ & 30 ps$^{-1}$ & $T_A$ & 2.00 ps & $\gamma$ & 0.00111 W$^{-1}$ \\			
			$l$ & 1.05 &&&&\\
			\hline
			$A_0$ & $25.2\,\sqrt{\mathrm{W}}$ &$\tau_0$& 143\,fs& $w_0$ & 0.182\,nJ\\
			\hline
		\end{tabular}
		\caption{{The values of parameters we use in Eqs.~(\ref{eq:hme}),~(\ref{eq:gain_sat},~(\ref{eq:sesam}), and~(\ref{eq:n_sesam}). The stationary pulse parameters $A_0$, $\tau_0$, and $w_0$ are obtained computationally and thus are separated from the rest. }\label{tab:params-sesam}}	
	\end{center}
\end{table}

\begin{figure}[h!]
	\begin{center}
		\includegraphics[width=0.65\linewidth]{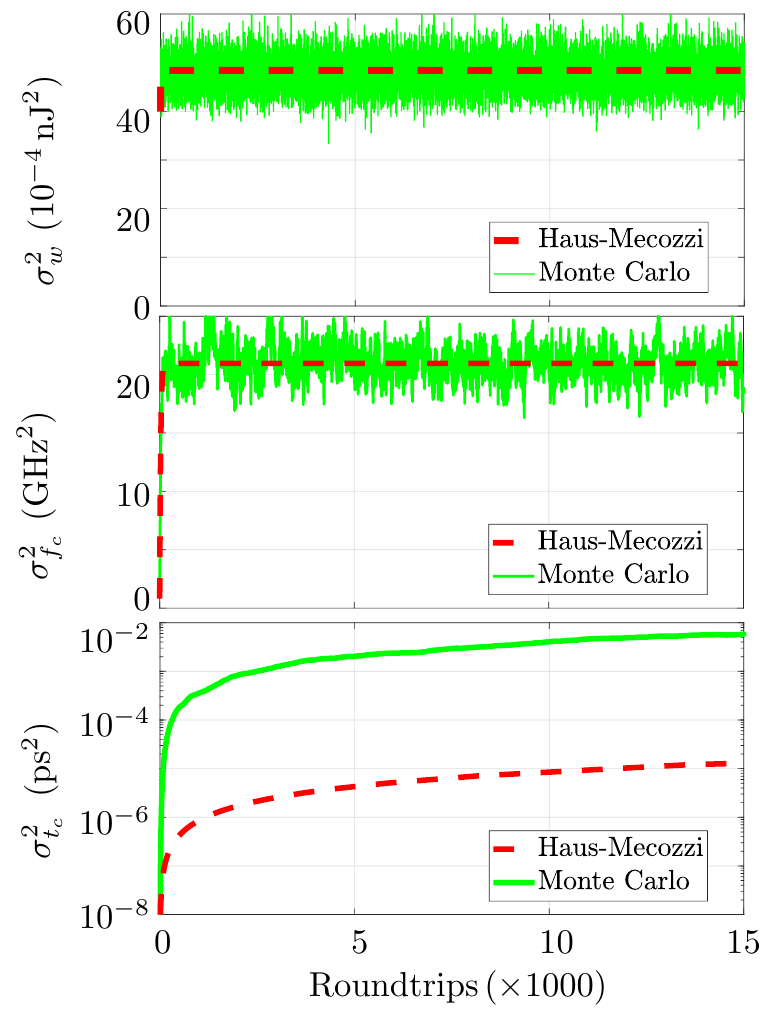}					
		\caption{Comparison between the Haus-Mecozzi and Monte Carlo methods for the SESAM fiber laser, where $\sigma^2_{w}(T)$, $\sigma^2_{f_c}(T)$, and $\sigma_{t_c}^2(T)$ are propagation-dependent variances of the pulse energy fluctuation $\Delta w$, central frequency $\Delta f_c$, and the central time $\Delta t_c$. We obtain the Haus-Mecozzi method results by substituting the computational stationary pulse solution parameters from Table~\ref{tab:params-sesam} into \quoeq{eq:variances-hmn}. \label{fig:propagate-SESAM}}
	\end{center}
\end{figure}

In Fig.~\ref{fig:propagate-SESAM}, we show the evolution of the variances of $\Delta w$, $\Delta f_c$, and $\Delta t_c$. 
To compute the variances using the Haus-Mecozzi method, we use the stationary pulse parameters that we obtained computationally by propagating the evolution equations. 
We see that the Haus-Mecozzi method provides a good prediction  for the variances of the energy $\Delta w$ and and the frequency $\Delta f_c$. 
However, the Haus-Mecozzi model underestimates the variance of the central time $\Delta t_c$ by a factor of $300$, as shown in Fig.~\ref{fig:propagate-SESAM}. 

\begin{figure}[h!]
	\begin{center}
		\includegraphics[width=0.7\linewidth]{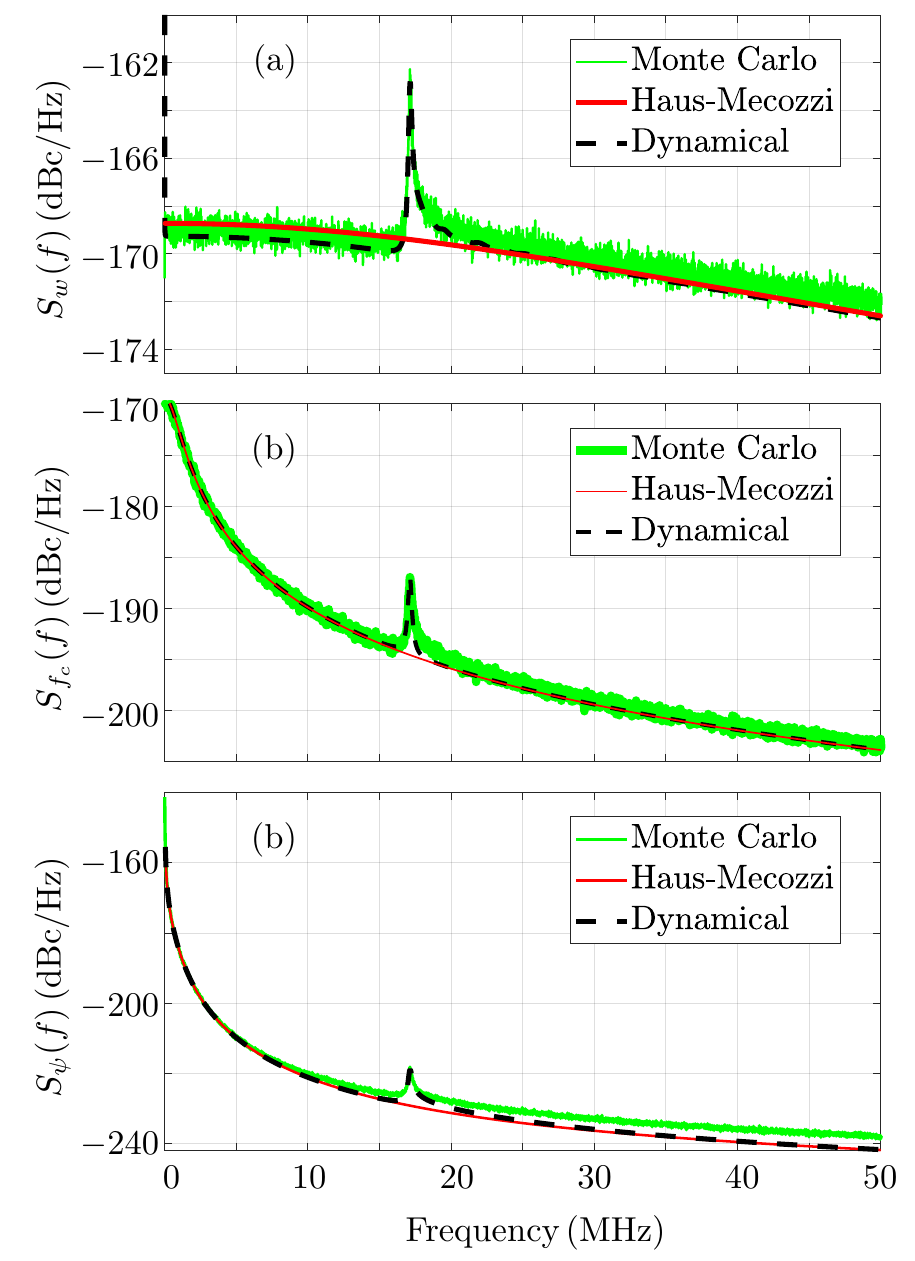}
		\caption{The power spectral density of (a) the energy jitter, (b) the frequency jitter, and (c) the timing phase jitter that we obtain from the Monte Carlo method, the Haus-Mecozzi method, and the dynamical method.\label{fig:psds-sesam}}
	\end{center}
\end{figure}

In Fig.~\ref{fig:psds-sesam}, we show the power spectral densities of $\Delta w$, $\Delta f_c$, and $\Delta t_c$ that we derived using these three methods. 
Both the Haus-Mecozzi method and the dynamical method yield good agreement for the background noise level with the Monte Carlo simulations. 
However, the Haus-Mecozzi method completely misses the sideband that is present in each of the power spectral densities. 
We have shown in prior work~\cite{Wang:2016a} that the output power spectrum of the SESAM fiber laser features a sideband that is located between 15\,MHz to 20\,MHz as the pump power changes. 
In the Monte Carlo simulations, the sideband appears in all three power spectral densities, as shown in Fig.~\ref{fig:psds-sesam}. 
The dynamical method is able to predict the height of the sidebands successfully. 
Hence, the dynamical method provides an accurate calculation of the noise levels for a wider group of modelocked lasers than does the Haus Mecozzi method. 

We observe that the Monte Carlo results consistently overestimate the noise level at higher frequencies, which is due to aliasing. 
We have defined the output signals of the laser cavity as a continuous-time random process.
However, in order to calculate the discrete-time Fourier transform, as in \quoeq{eq:psdEvaluate}, the output signal of the laser is recorded once per round trip, which sets an upper limit equal to the Nyquist frequency, which equals $1/(2T_R)=150$\,MHz. 
However, our noise source is wide-band. 
As a result, noise with frequencies higher than 150\,MHz will leak into our evaluation band and cause the evaluated noise level to rise. 
The Monte Carlo results will converge to the noise level that is obtained using the dynamical method when we record more times during one round trip, which increases the memory and post-processing load.

We again carry out a computational efficiency test, and we show the results in Table~\ref{tab:sesam-comp-test}.
Here, the Monte Carlo experiments are carried out using Matlab{\textregistered} and 512 cores on a cluster~\cite{maya_umbc}.
The CPUs are all quad-core Intel Nehalem X5560 processors (2.8\,GHz, 8\,MB cache) with 3\,GB per core on average. 
All nodes are running Red Hat Enterprise Linux 6.4. 
We propagate the pulse for 15000 rountrips, and we only save the data for the pulse parameters instead of the entire pulse. 
The entire computation requires about 20\,min and uses 256 computing cores. 
Each simulation takes more than 300\,MB on each computing core, and we saved 1.7\,MB of data on the hard drive. 

By comparison, the dynamical method is carried out on the same desk workstation as in Sec.~\ref{sec:hmn-compute}: a Dell{\textregistered} Precision Tower 7910 that uses an  Intel{\textregistered} Xeon(R) CPU E5-2630 v4, which includes 10 cores. 
From solving for the stationary solution to obtaining the power spectral density, the computational cost is less than 4\,min and uses very reasonable memory and storage. 
Again, the improvement in the computing efficiency is large. 
Compared to the Monte Carlo simulation method, the dynamical method requires only 1/1280 of the CPU time, 1/90 of the memory, and 1/3 of the storage space. 

\begin{table}
	\begin{center}
		\begin{tabular}{|l|c|c|c|c|}
			\hline
			Method & \# of cores & Time cost & Memory usage & Storage usage \\
			\hline
			256 runs & 256 & ~20 min & 314\,MB/process & 1.7\,MB/process\\
			\hline 
			Dynamical & 1 & {$<4$\,min} & ~900\,MB & 144\,MB\\
			\hline
		\end{tabular}
		\caption{Comparison of the computational efficiency of the Monte Carlo and dynamical methods for evaluating the noise performance of the SESAM modelocking model. We integrate the system for $2\times10^5$ round trips in each simulation run of the Monte Carlo method. \label{tab:sesam-comp-test}}
	\end{center}
\end{table}

\section{Conclusions\label{sec:conclude}}

Over the last three decades, short-pulse lasers---and more particularly passively modelocked lasers---have been the subject of continued experimental interest. 
Robust and low-noise passively modelocked lasers are the key component in frequency combs. 
As passively modelocked lasers have become more complex, the Haus-Mecozzi method has become increasingly inadequate to analyze the noise performance of these lasers. 
As one example, we studied a SESAM fiber laser and showed that this method greatly  underestimates the noise level. 
By contrast, Monte Carlo simulations can yield accurate results, and this method is intuitive and easy to implement. 
However, it requires large computing resources, which makes its use for parameter optimization difficult.

Based on dynamical systems theory, we have developed a dynamical method that makes it possible to calculate the noise levels accurately and rapidly. 
As we have shown in our examples, it is as accurate as Monte Carlo simulations, and is about three orders of magnitude faster computationally in our examples, while requiring less memory and storage. 
Therefore, this dynamical method is a powerful tool that can play a useful role in optimizing the design of short-pulse lasers. 

\section*{Appendix: Numerical Implementation}
When using modern-day scripting languages such as Matlab\textregistered\  and Python, 
it is more computationally efficient to carry out calculations using matrix operations. 
Here, we describe how to construct \quoeq{eq:psd-h} using matrix operations. 

We have discussed the computational discretization in Sec.~4 in~\cite{Wang:2014}.
We use $N$ to denote the number of points in the computational time window $T_w$;
we use $j$ to denote the row indices; and we use $k$ to denote the column indices. 

We begin by introducing the eigenvalue matrices $\mathsf{E}$ and $\hat{\mathsf{E}}$,
\begin{align}
\mathsf{E} = \begin{bmatrix}
|&|&\cdots&| \\
\vecd{e}_1 & \vecd{e}_2 & \cdots & \vecd{e}_{2N} \\
|&|&\cdots&|
\end{bmatrix}, \quad 
\hat{\mathsf{E}} = \begin{bmatrix}
|&|&\cdots&| \\
\hat{\vecd{e}}_1 & \hat{\vecd{e}}_2 & \cdots & \hat{\vecd{e}}_{2N} \\
|&|&\cdots&|
\end{bmatrix},
\end{align}
normalized so that $\hat{\mathsf{E}}^H\mathsf{E}\Delta t=\mathsf{I}_{2N}$, where $\mathsf{I}_{2N}$ is the identity matrix, which we use to define the matrix 
\begin{align}
\mathsf{D} =\paren{D\Delta t}\hat{\mathsf{E}}^H\hat{\mathsf{E}},
\end{align}
where $D$ is defined in \quoeqq{eq:defnoise}{eq:ase}.
We next define the matrix
\begin{align}
\mathsf{H} = \vecd{h}_e^*\vecd{h}_e^T, 
\end{align}
where $\vecd{h}_e^*$ is the element-wise complex conjugate of $\vecd{h}_e$ and  $\vecd{h}_e$ is defined in \quoeq{eq:calculate-h},
\begin{align}
\vecd{h}_e = \mathsf{E}^H\vecd{h}_x\Delta t.
\end{align}
Finally, we define the matrix 
\begin{align}
\mathsf{\Omega}(f) = \begin{bmatrix}
\mu_1 & \mu_1 & \cdots & \mu_1 \\
\mu_2 & \mu_2 & \cdots & \mu_2 \\
\vdots & \vdots & \vdots & \vdots \\
\mu_{2N} & \mu_{2N} & \cdots & \mu_{2N} \\
\end{bmatrix},
\end{align}
where $\mu_j=\lambda_j-2i\pi f$. 

We can now express \quoeq{eq:psd-h} in matrix form as
\begin{align}
S_{x}(f)= \sum_{j=1}^{2N}\sum_{{k}=1}^{2N} \mathsf{A}_{jk}(f),
\end{align}
in which the matrix $\mathsf{A}(f)$ is given by
\begin{align}
\mathsf{A}(f) =  \mathsf{D}\odot\mathsf{H}\oslash\sqpr{\mathsf{\Omega}(f)\odot\mathsf{\Omega}^H(f)},
\end{align}
where $\odot$ and $\oslash$ represent element-wise matrix multiplication and devision, respectively, and all matrices are $2N\times2N$ square matrices.

As an example, the Matlab{\textregistered} code that calculates the power spectral density of the timing phase jitter, shown in Fig.~\ref{fig:psds-sesam}(c), is available at \\ \hyperlink{http://photonics.umbc.edu/software.html}{http://photonics.umbc.edu/software.html}

\section*{Funding}
Aviation and Missile Research, Development, and Engineering Center (AMRDEC), Defense Advanced Research Projects Agency (DARPA) (W31P4Q-14-1-0002).

\section*{Acknowledgments}
We thank Patrick Sykes, Stefan Droste, Laura Sinclair, Ian Coddington, and Nathan Newbury for their support and useful discussions.  

\section*{References}
\bibliographystyle{elsarticle-num}
\bibliography{myrefs}		

\end{document}